\documentclass[preprint,preprintnumbers,amsmath,amssymb]{revtex4}

\usepackage{graphicx}% Include figure files
\usepackage{dcolumn}% Align table columns on decimal point
\usepackage{bm}% bold math

\begin{document}

%\preprint{PREPRINT L-1}

\title{Survival probability of large rapidity gaps in a QCD model with a
dynamical infrared mass scale}% Force line breaks with \\

\author{E.G.S. Luna$^{1,2}$}
%\altaffiliation[Also at ]{Physics Department, XYZ University.}%Lines break automatically
%or can be forced with \\
\affiliation{
$^{1}$Instituto de F\'{\i}sica Te\'orica,
UNESP, S\~ao Paulo State University, 01405-900,
S\~ao Paulo, SP, Brazil \\
$^{2}$Instituto de F\'{\i}sica Gleb Wataghin, Universidade Estadual de Campinas,
13083-970, Campinas, SP, Brazil}


\begin{abstract}
We compute the survival probability $\langle |S|^{2} \rangle$ of large rapidity gaps (LRG) in a QCD based eikonal
model with a dynamical gluon mass, where this dynamical infrared mass scale represents the onset of nonperturbative
contributions to the diffractive hadron-hadron scattering. Since rapidity gaps can occur in the case of Higgs
boson production via fusion of electroweak bosons, we focus on $WW\to H$ fusion processes and show that the
resulting $\langle |S|^{2} \rangle$ decreases with the increase of the energy of the incoming hadrons, in line
with the available experimental data for LRG. We obtain $\langle |S|^{2} \rangle = 27.6\pm7.8$ \%
($ 18.2\pm7.0$ \%) at Tevatron (CERN-LHC) energy for a dynamical gluon mass $m_{g}=400$ MeV.   
\end{abstract}

\maketitle

\section{Introduction}

The study of the survival probability $\langle |S|^{2} \rangle$ of large rapidity gaps (LRG) is currently a subject of
intense theoretical and experimental interest. Its importance lies in the fact that systematic analyses of
LRG open the possibility of extracting New Physics from hard diffractive processes. On the theoretical side, its
significance is due to the reliance of the $\langle |S|^{2} \rangle$ calculation on subtle QCD methods.

Rapidity gaps are defined as regions of angular phase space devoid of particles
\cite{dokshitzer,dokshitzer1,chehime,bjorken1,bjorken2}, and
at high-energy hadron-hadron collisions it has been suggested that
their observation may serve as a signature for production of Higgs bosons and other colour-singlet systems via
fusion of electroweak bosons \cite{dokshitzer,dokshitzer1,bjorken1,bjorken2,escobar1,fletcher1,block1}.
However, as pointed out by Bjorken \cite{bjorken1,bjorken2}, we are not able to distinguish between the
theoretically calculated rate of a large rapidity gap, $F_{gap}$, and the actual measured rate, $f_{gap}$. These
rates are related by the proportionality
relation $f_{gap} = \langle |S|^{2} \rangle \, F_{gap}$, where $\langle |S|^{2} \rangle$ is the so called ``survival
probability'' of a large rapidity gap: it gives the probability of survival of LRG and
not the probability for the production {\it and} survival of LRG, which is the quantity actually measured. More specifically,
the survival factor $\langle |S|^{2} \rangle$ gives the probability of a large rapidity gap not to be filled by debris
originated from the
rescattering of spectator partons ($\langle |S_{spec}|^{2} \rangle$), or from the emission of bremsstrahlung gluons from partons
($\langle |S_{brem}|^{2} \rangle$). Hence we can write
\begin{eqnarray}
\langle |S|^{2} \rangle = \langle |S_{brem} (\Delta y=|y_{1}-y_{2}|)|^{2} \rangle \langle |S_{spec}(s)|^{2} \rangle ,
\end{eqnarray}
where $y_{1}$ and $y_{2}$ are the rapidities of the jets (produced with large transverse momenta),
$\Delta y=|y_{1}-y_{2}|\gg 1$, and $s$ is the square of the total center-of-mass (CM)
energy. The factor $\langle |S_{brem} (\Delta y)|^{2} \rangle$, which depends on the value of the LRG as well as on the
kinematics of each specific process, can be calculated using perturbative QCD \cite{khoze4}. On the other hand, the survival
factor $\langle |S_{spec}|^{2} \rangle$ cannot be calculated perturbatively: it takes account of soft rescatterings in both the
initial and final state interactions. To calculate $\langle |S_{spec}|^{2} \rangle$ we need to obtain the probability of
the two incoming hadrons do not interact inelastically, i.e. the probability of large rapidity gaps do not be populated by
additional
hadrons produced from the rescattering of spectator degrees of freedom. As discussed by Gotsman, Levin and Maor some time
ago, this task is a difficult one since partons at long distances contribute to such a calculation \cite{gotsman1}. However,
the survival probability can be properly defined in the impact parameter space \cite{bjorken1,bjorken2}:
\begin{eqnarray}
\langle |S_{spec}|^{2} \rangle = \frac{\int d^{2}\vec{b} \, F(b,s)\, P(b,s)}{\int d^{2}\vec{b} \, F(b,s)} ,
\label{definit01}
\end{eqnarray}
where $b$ is the impact parameter, $F(b,s)$ is a factor related to the overlap of the parton densities in the colliding hadrons
in the transverse impact plane,
and $P(b,s)$ is the probability that the two incoming hadrons have not undergone a inelastic scattering at the parton
level. The above definition sets up the stage for carrying out a systematic computation of $\langle |S_{spec}|^{2} \rangle$,
since the probability $P(b,s)$ can be easily obtained, for example, from the QCD-inspired eikonal approach
\cite{gregores,pancheri1,godbole,durand}. In this picture the probability that
neither hadron is broken up in a collision at impact parameter $b$ is given by $P(b,s)=e^{-2\chi_{I}(b,s)}$, where
$\chi_{I}(b,s)$ is the imaginary part of the eikonal function. In QCD-inspired eikonal models the increase
of the total cross sections is associated with semihard scatterings of partons in the hadrons, and the high energy
dependence of the cross sections is driven mainly by gluon-gluon scattering processes. Nevertheless, the gluon-gluon
subprocess cross section is potentially divergent at small
transferred momenta, and the usual procedure to regulate this behaviour is the introduction of a purely
{\it ad hoc} parameter separating the perturbative from the non-perturbative QCD
region, like an infrared mass scale \cite{gregores,pancheri1}, or a cut-off at low transverse
momentum $p_{T}$ \cite{godbole,durand}.

Recently we introduced a QCD-based eikonal model where the {\it ad hoc} infrared mass scale was substituted by
a dynamical gluon mass one \cite{luna01}. One of the central advantages of the model is that it gives a precise
physical meaning for the quoted infrared scale. Furthermore, since the behaviour of the running coupling constant
is constrained by the value of dynamical gluon mass \cite{cornwall,ans}, the model also has a smaller number of
parameters than similar QCD models.

In this letter we perform a detailed computation of the survival probability in $pp$ and $\bar{p}p$ channels in the framework
of the QCD eikonal model with a gluon dynamical mass. We are concerned with the calculation of $\langle |S_{spec}|^{2}
\rangle$, the probability that LRG survive the soft rescattering of spectator partons, which we shall denote henceforth simply as
$\langle |S|^{2} \rangle$. In the next section we introduce the QCD eikonal model and address the question of calculating
the survival factor $\langle |S|^{2} \rangle$ from processes of Higgs boson production through $W$ fusion. The results are
presented in the Sec. III, where we provide a systematic study of $\langle |S|^{2} \rangle$ and its sensitivity to the infrared
mass scale. The conclusions are drawn in Sec. IV.

\section{Large rapidity gaps and the dynamical gluon mass}

It has been suggested that in hadron-hadron collisions the production of Higgs bosons can occur by means of the
fusion of gluons or electroweak bosons \cite{dokshitzer,dokshitzer1,chehime,bjorken1}. The two main Higgs production mechanisms are the
gluon fusion $gg\to H$ and the $W$ fusion $WW\to H$. In the gluon fusion process
each hadron emits a gluon (a colour octet), and the Higgs boson is coupled to each one through a fermion
loop. Since the Higgs boson couples to fermions according to their masses, its production cross section via gluon fusion is
dominated by top quark loops. In gluon fusion the hadrons remnants must exchange colour with each other
in order to become singlet states again. 

On the other hand, the $W$ fusion process has a different colour flow structure: since the incoming hadrons
(as well as the $W$ bosons) are colour singlets, when they emit a $W$ boson they remain as singlet states. Thus no
colour is exchanged between the scattered partons, and the two outgoing hadron remnant states are expected to be separated by a
central rapidity gap. As indicated in the previous section, the survival probability of these rapidity gaps can be
naturally defined in the impact parameter representation. In this formalism the inclusive differential Higgs boson production
cross section via $W$ fusion is given by
\begin{eqnarray}
\frac{d\sigma_{prod}}{d^{2}\vec{b}}=\sigma_{WW\to H}\, W(b;\mu_{W} ),
\label{eqnarrat01}
\end{eqnarray}
where $W(b;\mu_{W})$ is the overlap function at impact parameter space of the $W$ bosons. This function represents the effective
density of the overlapping $W$ boson distributions in the colliding hadrons. The cross section for producing the Higgs
boson and having a large rapidity gap is given by
\begin{eqnarray}
\frac{d\sigma_{LRG}}{d^{2}\vec{b}} &=& \sigma_{WW\to H}\, W(b;\mu_{W} )\, P(b,s),
\label{eqnarrat02}
\end{eqnarray}
where $P(b,s)$ is the probability that the two initial hadrons have not undergone a
inelastic scattering at the parton level. In QCD-inspired eikonal models this probability is given by
$P(b,s)=e^{-2\chi_{I}(b,s)}$, where the imaginary part $\chi_{I}(b,s)$ of the eikonal function receives contributions of
parton-parton interactions. Therefore, the factor $P(b,s)$
suppresses the contribution to the Higgs boson cross section where the two initial hadrons overlap and there is soft
rescatterings of the spectator partons. From the expressions (\ref{definit01}), (\ref{eqnarrat01}) and
(\ref{eqnarrat02}) we can write down the survival factor $\langle |S|^{2} \rangle$ for Higgs production via $W$ fusion:
\begin{eqnarray}
\langle |S|^{2} \rangle &=& \frac{\int d^{2}\vec{b} \, \sigma_{WW\to H}\, W(b;\mu_{W} )\,
e^{-2\chi_{I}(b,s)}}{\int d^{2}\vec{b} \,
\sigma_{WW\to H}\, W(b;\mu_{W} )} \nonumber \\
&=& \int d^{2}\vec{b} \, W(b;\mu_{W} )\, e^{-2\chi_{I}(b,s)} ,
\label{eqnarrat03}
\end{eqnarray}
where we have used the normalization condition $\int d^{2}\vec{b} \, W(b;\mu_{W} ) = 1$. In this letter we shall compute the
probability factor $P(b,s)=e^{-2\chi_{I}(b,s)}$ using a recently developed QCD eikonal model, where the onset of the
dominance of gluons in the interaction
of high-energy hadrons is managed by the dynamical gluon mass scale \cite{luna01}. The model, henceforth referred to as DGM
model, satisfies analyticity and unitarity
constraints. The latter is automatically satisfied in the eikonal representation, where the total cross section, the ratio
$\rho$ of the real to the imaginary part of the forward scattering amplitude, and the differential elastic scattering cross
section are given by
\begin{eqnarray}
\sigma_{tot}(s) = 4\pi \int_{_{0}}^{^{\infty}} \!\! b\, db\, [1-e^{-\chi_{_{I}}(b,s)}\cos \chi_{_{R}}(b,s)],
\label{degt1}
\end{eqnarray}
\begin{eqnarray}
\rho (s) = \frac{\textnormal{Re} \{ i \int b\, db\, [1-e^{i\chi (b,s)}]  \}}{\textnormal{Im} \{ i \int b\,
db\, [1-e^{i\chi (b,s)}]  \}},
\label{degthyj1}
\end{eqnarray}
and
\begin{eqnarray}
\frac{d\sigma_{el}}{dt}(s,t)=\frac{1}{2\pi}\, \left| \int b\, db\, [1-e^{i\chi (b,s)}]\, J_{0}(qb) \right|^2 ,
\label{degthyj3}
\end{eqnarray}
respectively, where $s$ is the square of the total CM energy, $J_{0}(x)$ is the Bessel function of the first kind,
and $\chi(b,s)=\chi_{_{R}}(b,s)+i\chi_{_{I}}(b,s)$ is the (complex) eikonal function. In the DGM model the eikonal function is
written as a combination of an even and odd eikonal terms related by crossing symmetry. In terms of the proton-proton ($pp$) and
antiproton-proton ($\bar{p}p$) scatterings, this combination reads
$\chi_{pp}^{\bar{p}p}(b,s) = \chi^{+} (b,s) \pm \chi^{-} (b,s)$. 
The even eikonal is written as the sum of gluon-gluon, quark-gluon, and quark-quark contributions:
\begin{eqnarray}
\chi^{+}(b,s) &=& \chi_{qq} (b,s) +\chi_{qg} (b,s) + \chi_{gg} (b,s) \nonumber \\
&=& i[\sigma_{qq}(s) W(b;\mu_{qq}) + \sigma_{qg}(s) W(b;\mu_{qg})+ \sigma_{gg}(s) W(b;\mu_{gg})] ,
\label{final4}
\end{eqnarray}
where $W(b;\mu)$ is the overlap
function at impact parameter space and $\sigma_{ij}(s)$ are the elementary subprocess cross sections of colliding quarks and
gluons ($i,j=q,g$). The overlap function, normalized so that $\int d^{2}\vec{b}\, W(b;\mu)=1$, is associated with the Fourier
transform of a dipole form factor,
\begin{eqnarray}
W(b;\mu) = \frac{\mu^2}{96\pi}\, (\mu b)^3 \, K_{3}(\mu b),
\label{dipole10}
\end{eqnarray}
where $K_{3}(x)$ is the modified Bessel function of second kind. The odd eikonal $\chi^{-}(b,s)$, that
accounts for the difference between $pp$ and $\bar{p}p$ channels, is parametrized as
\begin{eqnarray}
\chi^{-} (b,s) = C^{-}\, \Sigma \, \frac{m_{g}}{\sqrt{s}} \, e^{i\pi /4}\, 
W(b;\mu^{-}),
\label{oddeik}
\end{eqnarray}
where $m_{g}$ is the dynamical gluon mass and the parameters $C^{-}$ and $\mu^{-}$ are constants to be
fitted. The factor $\Sigma$ is defined as
\begin{eqnarray}
\Sigma = \frac{9\pi \bar{\alpha}_{s}^{2}(0)}{m_{g}^{2}},
\end{eqnarray}
with the dynamical coupling constant
$\bar{\alpha}_{s}$ set at its frozen infrared value. The origin of the dynamical gluon mass and the
frozen coupling constant can be traced back to the early work of Cornwall \cite{cornwall}, and the formal
expressions of these quantities can be seen in Ref. \cite{luna01}. 

The eikonal functions $\chi_{qq} (b,s)$ and $\chi_{qg} (b,s)$, needed to describe the lower-energy forward data, are simply
parametrized with terms dictated by the Regge phenomenology:
\begin{eqnarray}
\chi_{qq}(b,s) = i \, \Sigma \, A \,
\frac{m_{g}}{\sqrt{s}} \, W(b;\mu_{qq}),
\label{mdg1}
\end{eqnarray}
\begin{eqnarray}
\chi_{qg}(b,s) = i \, \Sigma \left[ A^{\prime} + B^{\prime} \ln \left( \frac{s}{m_{g}^{2}} \right) \right] \,
W(b;\sqrt{\mu_{qq}\mu_{gg}}),
\label{mdg2}
\end{eqnarray}
where $A$, $A^{\prime}$, $B^{\prime}$, $\mu_{qq}$ and $\mu_{gg}$ are fitting parameters. The gluon-gluon eikonal
contribution, that dominates at high energy and determines the asymptotic behaviour of the total cross sections, is written as
$\chi_{gg}(b,s)\equiv \sigma_{gg}^{D\!PT}(s)W(b; \mu_{gg})$, where
\begin{eqnarray}
\sigma_{gg}^{D\!PT}(s) = C^{\prime} \int_{4m_{g}^{2}/s}^{1} d\tau \,F_{gg}(\tau)\,
\hat{\sigma}^{D\!PT}_{gg} (\hat{s}) .
\label{sloh1}
\end{eqnarray}

Here $F_{gg}(\tau)$ is the convoluted structure function for pair $gg$, $\hat{\sigma}^{D\!PT}_{gg}(\hat{s})$ is
the subprocess cross section and $C^{\prime}$ is a fitting parameter. In the
above expression it is introduced the energy threshold $\hat{s}\geq 4m_{g}^{2}$ for the final state gluons,
assuming that these are screened gluons \cite{cornwall2}. The structure function $F_{gg}(\tau)$ is given by
\begin{eqnarray}
F_{gg}(\tau)=[g\otimes g](\tau)=\int_{\tau}^{1} \frac{dx}{x}\, g(x)\,
g\left( \frac{\tau}{x}\right),
\end{eqnarray}
where $g(x)$ is the gluon distribution function, adopted as
\begin{eqnarray}
g(x) = N_{g} \, \frac{(1-x)^5}{x^{J}},
\label{distgf}
\end{eqnarray}
where $J=1+\epsilon$ and $N_{g}=\frac{1}{240}(6-\epsilon)(5-\epsilon)...(1-\epsilon)$. The correct analyticity properties of the
model amplitudes is ensured by substituting $s\to se^{-i\pi/2}$ throughout Eqs. (\ref{mdg1}), (\ref{mdg2}) and (\ref{sloh1}).

In the expression (\ref{sloh1}) the gluon-gluon subprocess cross section $\hat{\sigma}^{D\!PT}_{gg}(\hat{s})$ is calculated
using a procedure dictated by the dynamical perturbation theory (DPT) \cite{pagels}: amplitudes that do not vanish to all
orders of perturbation theory are given by their free-field values, whereas amplitudes that vanish in all orders
in perturbation theory as $\propto \exp{(-1/g^2)}$ ($g$ is the coupling constant) are retained at lowest
order. In our case this means that the effects of the dynamical gluon mass in the propagators and vertices
are retained, and the sum of polarizations is performed for massless (free-field) gluons. As a result, since the dynamical
masses go to zero at large momenta, the elementary cross sections of perturbative QCD in the high-energy limit are recovered.
Other details of the calculation can be seen in Ref. \cite{luna01}.

According to the expression (\ref{eqnarrat03}), the final step in order to calculate $\langle |S|^{2} \rangle$ is
to determine the overlap function $W(b;\mu_{W})$ of the electroweak bosons. It is worth mentioning that the survival factor
$\langle |S|^{2} \rangle$ depends on the nature of the colour-singlet exchange which generates the gap as well as on the
distributions of partons inside the proton in impact parameter space \cite{gotsman1,khoze1,khoze2,khoze3,gotsman2}.
We simply assume that the distribution of $W$ bosons in impact parameter space in the hadron
is the same as for the quarks. In this way, we can finally write down a phenomenologically useful expression to
the survival factor $\langle |S|^{2} \rangle$:
\begin{eqnarray}
\langle |S|^{2} \rangle = 2\pi \int_{0}^{\infty} b\, db \, W(b;\mu_{qq} )\, e^{-2\chi_{I}(b,s)} .
\label{eqnarrat09}
\end{eqnarray}

In the above expression the inverse size (in impact parameter) $\mu_{qq}$ is the same as in the
expression (\ref{mdg1}). Its value, as well as the value of the remaining fitting parameters, is determined from global fits to
$pp$ and $\bar{p}p$ forward scattering data, as discussed in the next section.

\section{Results}

In order to determine the exponential damping factor $e^{-2\chi_{I}(b,s)}$ of Eq. (\ref{eqnarrat09}) and produce a consistent
estimate of $\langle |S|^{2} \rangle$, we carry out global fits to the elastic differential scattering cross section for
$\bar{p}p$ at $\sqrt{s}=1.8$ TeV and to all high-energy forward $pp$ and $\bar{p}p$ scattering data
above $\sqrt{s}=15$ GeV. This energy threshold is the same one used in the estimate of $\langle |S|^{2} \rangle$
through the analysis of $pp$ and $\bar{p}p$ scattering carried out by Block and Halzen using a previous QCD-inspired
model \cite{block1}.
The forward data sets include the total cross section ($\sigma_{tot}$) and the ratio of the real to imaginary part of the forward
scattering amplitude ($\rho$). We use the data sets compiled and analyzed by the Particle Data Group \cite{eidelman}, with
the statistic and systematic errors added in quadrature. The input values of the $m_{g}$ have been chosen
to lie in the interval $[350,650]$ MeV, as suggested by the
value $m_{g}=400^{+350}_ {-100}$ MeV obtained in a previous analysis of the $pp$ and $\bar{p}p$ channels via the DGM
model \cite{luna01}. This input dynamical gluon mass range is also supported by recent studies on the $\gamma p$ photoproduction
and the hadronic $\gamma \gamma$ total cross sections \cite{luna02}, and on the behaviour of the gluon distribution
function at small $x$ \cite{luna03}. In all the fits performed in this letter we use a $\chi^{2}$ fitting
procedure, assuming an interval
$\chi^{2}-\chi^{2}_{min}$ corresponding, in the case of normal errors, to the projection of the
$\chi^{2}$ hypersurface containing 90\% of probability. In the case of the DGM model (8 fitting parameters)
this corresponds to the interval $\chi^{2}-\chi^{2}_{min}=13.36$.

The $\chi^{2}/DOF$ values obtained in the global fits are relatively low, as show in Table I. These results (for 168 degrees
of freedom) indicate the excellence of the fits and show that the DGM model naturally accommodates all the data sets used
in the fitting procedure. In Table I we have included the
values of the $\mu_{W}(\equiv \mu_{qq})$ parameter, which determines the spatial distribution of the $W$ bosons at the impact
parameter $b$. We can observe a small dependence of their values on the dynamical gluon mass: the greater the input scale
$m_{g}$, the smaller the inverse size $\mu_{W}$.

The sensitivity of the survival probability $\langle |S|^{2} \rangle$ (for $pp$ collisions) to the gluon dynamical mass is
shown in Figure 1 for some CM energies. We note a slow increase of their values with the  
dynamical gluon mass and a fast decrease with the CM energies. As shown in Figure 2, where we have also
plotted the exponential damping factor, this behaviour is related to the energy dependence of the
imaginary part of the eikonal: $\chi_{I}(b,s)$ grows with the energy and hence suppresses the integral
(\ref{eqnarrat09}). In Table II we list our results for the survival factor $\langle |S|^{2} \rangle$ for some values of
the proton-proton energy, and compare with other calculations in the literature, where
$\langle |S|^{2} \rangle_{DGM1}$ and $\langle |S|^{2} \rangle_{DGM2}$
denote the results obtained by setting the mass infrared scale at $m_{g}=400$ and 600 MeV, respectively. The last value is
the same one adopted by Block and Halzen with respect to the {\it ad hoc} mass scale $m_{0}$ \cite{block1}. We see that the
DGM results are systematically larger than the Block-Halzen ones (denoted by $\langle |S|^{2} \rangle_{BH}$), but the large
statistical errors resulting from our choice for the confidence region of the parameters (CL=90 \%) indicate a reasonable
compatibility between the $\langle |S|^{2} \rangle_{DGM1}$ and $\langle |S|^{2} \rangle_{BH}$ results at higher energies.

The $\langle |S|^{2} \rangle_{DGM2}$ results at $\sqrt{s}=1.8$ and 16 TeV are in line with the
$\langle |S|^{2} \rangle_{GLM1}$ ones, obtained by Gotsman and collaborators using a Regge pole model \cite{gotsman3}. In their
approach they use an eikonalized version of the Donnachie-Landshoff model in order to satisfy unitarity \cite{cudell2}. The
authors argue that their relatively large values for $\langle |S|^{2} \rangle$ can be reduced by an appropriate change in some
parameters included in their Gaussian approximation for $F(b,s)$ and $P(b,s)$ factors. We hasten to emphasise that the
$\langle |S|^{2} \rangle_{DGM1}$ results have been obtained using the preferred statistical value of the dynamical gluon
mass for $pp$ and $\bar{p}p$ scattering, namely $m_{g}=400$ GeV \cite{luna01}. In this case we believe that an eventual change
in the parameters of the GLM1 model may reduce their results in such a way to be compatible with the DGM1 ones.  

The $\langle |S|^{2} \rangle_{KMR}$ results have been obtained by Khoze and collaborators using a two-channel eikonal
model which embodies pion-loop insertions in the Pomeron trajectory, diffractive dissociation, and rescattering
effects \cite{khoze1}. The authors have calculated the survival probability $\langle |S|^{2} \rangle$ in single, central and
double diffractive processes at several energies, assuming that the spatial distribution in the parameter space is controlled by
the slope $b$ of the Pomeron-proton vertex. We show the $\langle |S|^{2} \rangle_{KMR}$ results for
double diffractive processes with $2b=5.5$ GeV$^{2}$, which corresponds to the slope of the electromagnetic proton form factor.
These results are compatible with the DGM ones, in particular with the results taking into account $m_{g}=400$ MeV, the optimum
value for the dynamical gluon mass in $pp$ and $\bar{p}p$ diffractive scattering \cite{luna01}.

\section{Conclusions}

In this letter we have calculated the survival probability $\langle |S|^{2} \rangle$ of large rapidity gaps by means of
an eikonal QCD model with a dynamical gluon mass. Since rapidity gaps can occur from production of Higgs boson via fusion of
electroweak bosons, we have focused on $WW\to H$ fusion processes. The eikonal function have been determined
by fitting $pp$ and $\bar{p}p$ accelerator scattering data. Owing to the quality of the global fits, the DGM model allows us
to describe successfully the $\bar{p}p$ differential cross section at $\sqrt{s}=1.8$ TeV, as well as the forward scattering
quantities $\sigma_{tot}^{\bar{p}p,pp}$ and $\rho^{\bar{p}p,pp}$, in excellent agreement with the available experimental
data. These results show that the DGM model is well suited for the prediction of the survival probability of LRG at higher
energies, in particular for $\langle |S|^{2} \rangle$ one at the CERN-LHC energy.

In Table II we list our results for $\langle |S|^{2} \rangle$ and notice
that their values decrease with the increase of the energy of the incoming protons. This
behaviour, in line with results for LRG dijet production at the Tevatron \cite{D0,CDF6}, is a direct consequence of the energy
dependence of the imaginary part of the eikonal, that grows with the energy. A strong dependence of $\langle |S|^{2} \rangle $
on the dynamical gluon mass $m_{g}$ emerges from our calculations, as shown in Figure 1. This scale dependence arises as follows:
the dynamical gluon mass affects strongly the behaviour of the gluon-gluon subprocess cross section
$\hat{\sigma}^{D\!PT}_{gg}(\hat{s})$, which dominates at high energy and determines the asymptotic behaviour of the $pp$ and
$\bar{p}p$ total cross section. Hence the procedure consisting of global fits to diffractive $pp$ and
$\bar{p}p$ data in order to determine $m_{g}$ is well justified, and the value $m_{g}=400^{+350}_ {-100}$ MeV obtained in Ref.
\cite{luna01} via the DGM model is a suitable one.

Our estimates for the survival probability of large rapidity gaps using a QCD based eikonal model with a dynamical gluon mass
are, within the errors, compatible with estimates obtained using other eikonal models. In particular, our estimates are
close to the ones obtained by Khoze {\it et al.} using a two-channel model, and to the ones obtained by Block and
Halzen using a similar QCD-inspired approach. Owing to the interval [300,750] MeV inferred from the optimal value
of $m_{g}$ discussed above, there is room for smaller values
of the survival factor in DGM model. For example, a mass scale $m_{g}\sim 300$ MeV gives a survival factor
$\langle |S|^{2} \rangle \sim 15.3$ \% at LHC, very close to the central value obtained via the KMR model.

However, we call attention to the fact that all these estimates are model
dependent, despite their apparent agreement. For
example, the $\langle |S|^{2} \rangle_{KMR}$ results for other values of $2b$ and for central and single diffractive processes
do not agree with ours \cite{khoze1}. The same is expected in the $\langle |S|^{2} \rangle_{BH}$ results for other choices of
the mass scale $m_{0}$.

In summary, there is a strong dependence in the size of the survival probabilities and in their energy dependence on specific
models for the rise of total hadronic cross section. More specifically, the survival factor $\langle |S|^{2} \rangle$
depends on the dynamics of the whole diffractive part of the scattering matrix as well as the nature of the colour-singlet
exchange which generates the gap. From the experimental viewpoint, it
is known that the survival factor $\langle |S|^{2} \rangle$ in the case of Higgs production via $WW\to H$ fusion processes can
be monitored by observing the closely related central production of a $Z$ boson with the same jet and rapidity gap
configuration \cite{chehime2}. More recently, this idea has been developed further by considering the decays of both Higgs
and $Z$ bosons into $\bar{b}b$ pairs \cite{khoze4}. This allows to gauge Higgs weak boson fusion production at the LHC and
permits to observe a light Higgs boson via its dominant $H\to \bar{b}b$ decay mode in addiction to the usually discussed
$\tau \tau$ and $W W^{*}$ channels. This option would permit to reduce the theoretical uncertainty in the rate of Higgs central
production events with rapidity gaps.

The success of the QCD-based eikonal model with a dynamical gluon mass in reproducing diffractive scattering data, over a large
energy range, shows that such a model provides a reliable estimate of the survival probability $\langle |S|^{2} \rangle$ as a
function of energy in the case of $pp$ and $\bar{p}p$ channels. The study of the survival factor $\langle |S|^{2} \rangle$ is
interesting in its own right since they enables us to increase
our understanding of some features of hadronic interactions, and may provide an useful tool to probe physics beyond the Standard
Model.

\begin{acknowledgments}
We would like to thank C.O. Escobar, V.A. Khoze, M.J. Menon, and A.A. Natale for useful comments. This research was supported by
the Conselho Nacional de Desenvolvimento Cient\'{\i}fico e Tecnol\'ogico-CNPq under contract 151360/2004-9.
\end{acknowledgments}

\newpage

\begin{table*}
\caption{The $\mu_{W}$ parameter as a function of the dynamical gluon mass $m_{g}$. The $\chi^{2}/DOF$ values
resulting from the global fits are obtained for 168 degrees of freedom.}
\begin{ruledtabular}
\begin{tabular}{ccc}
$m_{g}$ [GeV] & $\mu_{W}$ [GeV] & $\chi^{2}/DOF$ \\
\hline
350 & 0.8308$\pm$0.1394 & 1.043 \\
400 & 0.8091$\pm$0.1410 & 1.022 \\
450 & 0.7848$\pm$0.1411 & 1.010 \\
500 & 0.7823$\pm$0.1392 & 1.009 \\
550 & 0.7227$\pm$0.1356 & 1.000 \\
600 & 0.7254$\pm$0.1333 & 1.001 \\
650 & 0.7025$\pm$0.1305 & 0.999 \\
% 700 & 0.6976$\pm$0.1258 & 1.000 \\
% 750 & 0.7009$\pm$0.1269 & 1.000 \\
\end{tabular}
\end{ruledtabular}
\end{table*}

\begin{table*}
\caption{The survival probability $\langle |S|^{2}\rangle$ (in \%) for $pp$ collisions in different models.}
\begin{ruledtabular}
\begin{tabular}{cccccc}
$\sqrt{s}$ [GeV] & $\langle |S|^{2}\rangle_{DGM1}$ & $\langle |S|^{2}\rangle_{DGM2}$ & 
$\langle |S|^{2}\rangle_{BH}$ & $\langle |S|^{2}\rangle_{GLM1}$ & $\langle |S|^{2}\rangle_{KMR}$ \\
% (in GeV) & (in \%) & (in \%) & (in \%) & (in \%) & (in \%) \\
\hline
63 & 45.4$\pm$8.4 & 50.9$\pm$9.3 & 37.5$\pm$0.9 & - & - \\
% 200 & 39.8$\pm$8.3 & 45.2$\pm$0.8 & - & - & - \\
546 & 34.2$\pm$8.1 & 39.4$\pm$8.9 & 26.8$\pm$0.5 & - & 26.0 \\
630 & 33.4$\pm$8.1 & 38.6$\pm$8.9 & 26.0$\pm$0.5 & - & - \\
1800 & 27.6$\pm$7.8 & 32.6$\pm$8.8 & 20.8$\pm$0.3 & 32.6 & 21.0 \\
% 1960 & 27.2$\pm$0.8 & 32.$\pm$0.8 & - & - & - \\
14000 & 18.2$\pm$7.0 & 22.8$\pm$8.3 & 12.6$\pm$0.06 & - & 15.0 \\
16000 & 17.7$\pm$6.9 & 22.6$\pm$8.2 & - & 22.1 & - \\
\end{tabular}
\end{ruledtabular}
\end{table*}

\begin{figure}
\label{figpaperluna01}
\vspace{2.0cm}
\begin{center}
%\vspace{-0.6cm}
\includegraphics[height=.60\textheight]{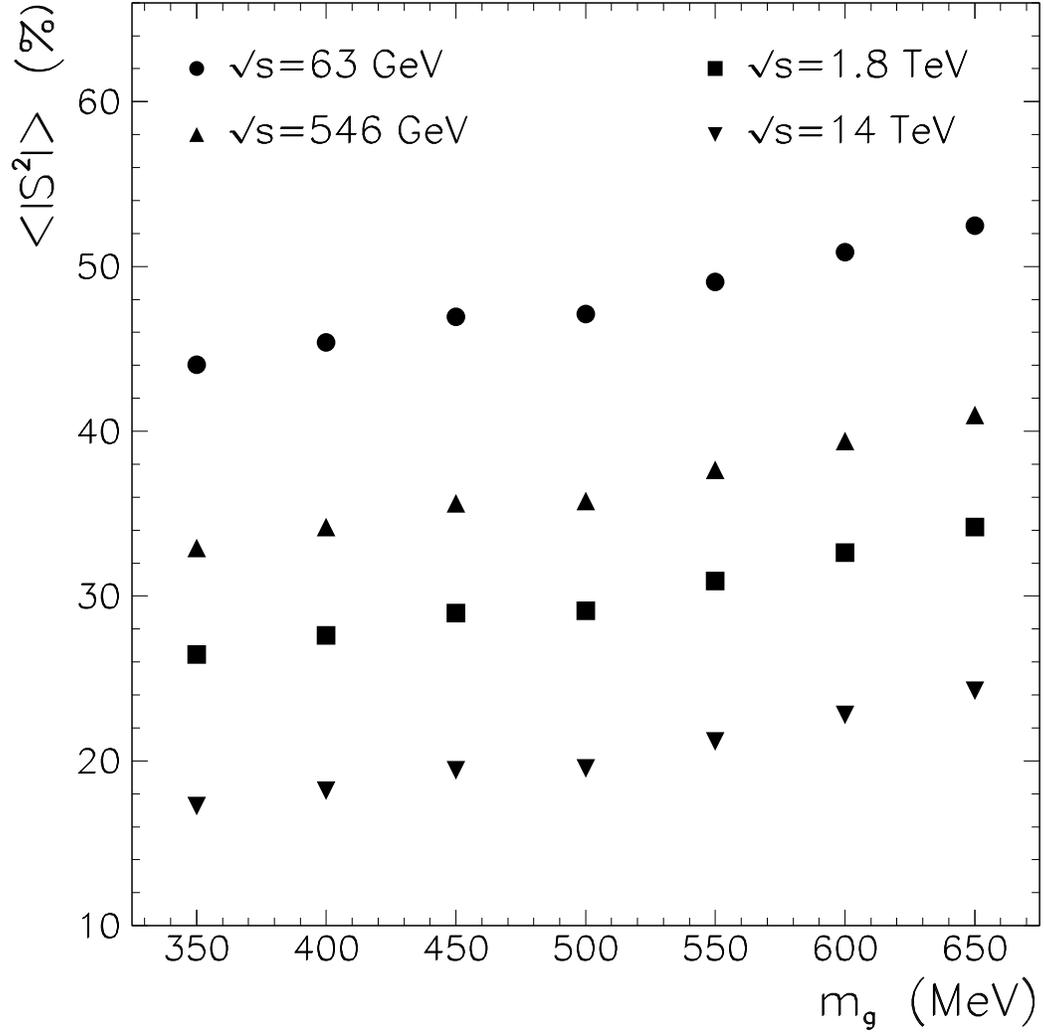}
\caption{The survival probability $\langle |S|^{2}\rangle$ (central values) as a function of the dynamical gluon mass $m_{g}$.}
\end{center}
\end{figure}

\begin{figure}
\label{figpaperluna02}
\vspace{2.0cm}
\begin{center}
%\vspace{-0.6cm}
\includegraphics[height=.60\textheight]{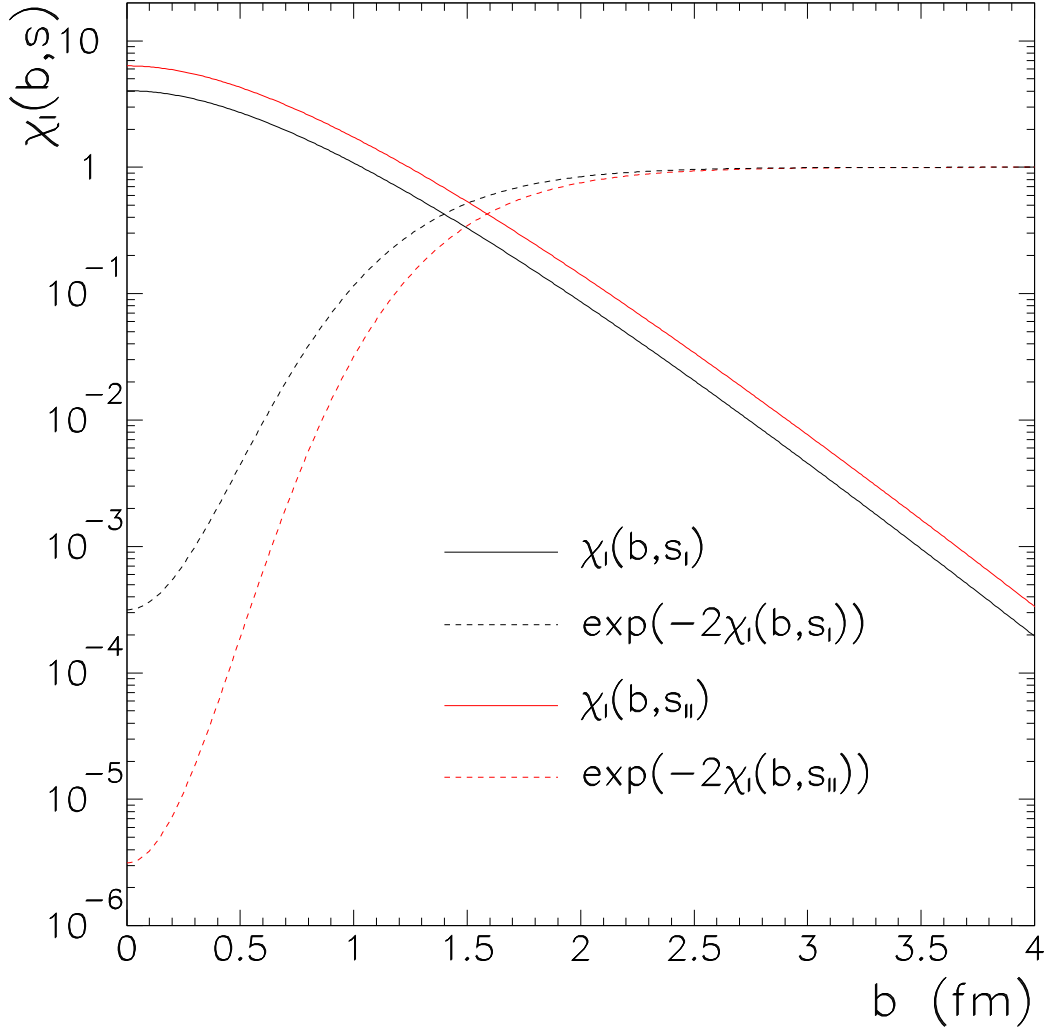}
\caption{The imaginary part $\chi_{I}(b,s)$ of the eikonal and the exponential factor $e^{-2\chi_{I}(b,s)}$
for $pp$ collisions as a function of the impact parameter $b$, where $\sqrt{s_{_{I}}}=1.8$ TeV and
$\sqrt{s_{_{II}}}=14$ TeV. The dynamical gluon mass scale was set to $m_{g}=400$ MeV.}
\end{center}
\end{figure}

\end{document}